# Possible violation of the optical theorem in LHC experiments

M Kupczynski

Département de l'Informatique, UQO, Case postale 1250, succursale Hull, Gatineau. QC, Canada J8X 3X 7

marian.kupczynski@uqo.ca

**Abstract**. The optical theorem allowing the determination of the total cross section for a hadron-hadron scattering from the imaginary part of the forward elastic scattering amplitude is believed to be an unavoidable consequence of the conservation of probability and of the unitary S matrix. This is a fundamental theorem which contains not directly measurable imaginary part of the forward elastic scattering amplitude. The impossibility of scattering phenomena without the elastic channel is considered to be a part of the quantum magic. However if one takes seriously the idea that the hadrons are extended particles one may define a unitary S matrix such that one cannot prove the optical theorem. Moreover data violating the optical theorem do exist but they are not conclusive due to the uncertainties related to the extrapolation of the differential elastic cross-section to the forward direction. These results were published several years ago but they were forgotten. In this paper we will recall these results in an understandable way and we will give the additional arguments why the optical theorem can be violated in high energy strong interaction scattering and why it should be tested and not simply used as a tool in LHC experiments.

## 1. Introduction.

Using a classical mechanics (CM) Rutherford described with success a scattering of alpha particles on a thin foil of gold as a scattering of point like charged particles by point-like positively charged particles in the target giving a clear indication that each atom contains a positively charged and heavy nucleus attracting negatively charged point-like electrons. There are no point-like physical objects in Nature but the point-like approximation (PLA), used often in CM with success for example to describe the motion of planets around the Sun, seemed to be justified because diameters of nuclei were of the order of $10^{-15}$ m and diameters of atoms of the order of $10^{-10}$ m.

In a quantum mechanics (QM) elastic scattering phenomena are described in a center of mass frame (CMS) as a scattering of some complex valued probability wave on a scattering center. For Coulomb elastic scattering one obtains the same formula as the formula found by Rutherford.

In QM for a scattering by short range potentials one obtains a startling relation, called the *optical theorem* (OT), between the imaginary part of the elastic forward scattering amplitude and the total elastic cross section.

The proof of OT can be generalized to cover various non-elastic scattering phenomena in atomic and molecular physics [1-3]. OT is also proven in the relativistic $S$ matrix theory [4-6]. The conservation of probability implies the *unitarity* of $S$ matrix. Since OT is proven using explicitly the *unitarity equation* it is considered to be a fundamental law and it is used as an important tool in various theoretical models and in the analysis of the experimental data. In particular it is used as a constraint in the maximum likelihood fits to elastic differential cross-section data.

Many years ago trying to understand why, using S matrix, one could not have high energy scattering without an elastic channel we succeeded to prove [7-10] that it was possible to construct a unitary $S$ matrix without OT. Simply instead of defining a scattering operator $T$ using the decomposition $S=I+iT$ we defined a unitary scattering operator $\tilde{S}$ by a formula $I \oplus \tilde{S}$ where $\tilde{S}$ was acting only on two particle initial state vectors with impact parameters smaller than the effective range of strong interactions. Our definition was consistent with an intuitive picture of colliding extended hadrons which in order to interact strongly have to hit each other.

The decomposition $S=I+iT$ is made in analogy to wave phenomena. However following Bohr we can say that we are dealing in hadron-hadron scattering rather with particle-like phenomena in which particle beams are prepared using the laws of relativistic classical mechanics and classical electrodynamics. These laws fail to describe what happens during a strong interaction scattering and we have to use some other theoretical models in order to make predictions about the type of particles produced and about the branching ratios of different reaction channels.

It is obvious that hadrons cannot be treated as point-like particles when we are studying a deep inelastic electron-hadron scattering or a strong hadron-hadron scattering. One can expect that completely new physics is needed to explain these phenomena. This is why in order to explain multiple hadron production observed in ultra-high energy cosmic rays (UHECR) data several quite successful statistical and thermo dynamical models of extended hadrons were proposed in the past by Fermi [11], Hagedorn [12], Frautschi [13] and several other researchers.

The extensive review of these models was given by Feinberg [14]. Various improved variants of these models continue to be used [15]. Other extended hadron models explaining the confinement of quarks in hadrons have been constructed: so called MIT bag model [16], chiral bag model [17] and their modifications.

In spite of the fact that the use of PLA leads to infinite self-energies in classical electrodynamics and to infinities in quantum field theory (QFT) requiring the renormalization PLA is still maintained for parton-parton interactions in so called standard model (SM) which is successfully used in the elementary particle physics.

In SM the confinement of quarks is taken for granted and one is using the parton-phenomenology together with the renormalizable QCD theory [18-20]. A hadron is described as composed of a number of point-like constituents called "*partons*": colored quarks and gluons. These point-like "free constituents" from two colliding hadrons interact instantaneously and incoherently producing in general several quark-antiquark pairs and gluons which recombine in the process of "*hadronization*" to form the final particles. The *parton* interactions are described by QCD and experimentally determined generalized *parton distribution functions* (GPD's) [18] are available worldwide.

The comparison of SM with the experimental data is a difficult task requiring the use of many free parameters, various phenomenological inputs and Monte Carlo simulation of events [19]. Moreover perturbative calculations of *parton-parton* interactions break down for small momentum transfers essential for the study of the elastic scattering [20].

Therefore to study the elastic cross sections single and multichannel *eikonal models* or/and *Regge phenomenology* are used with mixed successes [21-23]. In all these models and in the data analysis OT is used as an important constraint. All recent reported values for the total cross-sections [24-26] have been obtained using OT.

One can only conclude that physicists using SM both theoreticians and experimentalists are unaware of the fact that the violation of OT can be made consistent with the unitary *S* matrix. For this reason we review in some detail and in different way our forgotten results [7-9].

After proving that OT could be violated we inspected various elastic scattering data and we found the confirmation of our doubts [27]. However the only conclusion we could arrived at was that a reliable direct test of OT was impossible since for the same set of data one could get very good fits to the differential elastic cross sections: one drastically violating the OT and another consistent with it [9, 27].

Therefore we decided to search for other implications of our model according to which initial two-particle states were mixed quantum states with respect to some distance sensitive quantum number such as the impact parameter. Pure and mixed statistical ensembles have different properties [28]. Any sub-ensemble of a pure ensemble has the same properties as the initial ensemble. For a mixed ensemble one can find sub-ensembles with different properties. To be able to detect such differences one should search for some fine structure in the experimental data [28] using non-parametric statistical compatibility tests which we called *purity tests*[29-32].

However if one reanalyzes Tevatron , LHC and UHECR data without using OT one will find probably more  indications that OT may be violated.  There are several problems with the description of the elastic pp and other scattering data [20, 22, 23] which we are going to discuss in a subsequent more technical paper. If a model  using some theoretical assumption (such as OT) and several free parameters compares reasonably well with the experimental data [22] it does not prove that this particular theoretical assumption used is correct [9,28].

One can expect that with growing total collision energies up to 14 TeV, presently available at LHC, inelastic scattering channels will progressively be favored and the elastic scattering due to strong interactions will gradually be suppressed violating various bounds deduced using OT.

If hadrons are really extended particles they have to collide to interact strongly and then all allowed inelastic channels are open. If they are far enough and miss each other there is no strong interaction. It still could be the elastic channel open due to the Coulomb scattering for larger values of impact parameters but not due to strong interactions [7-10].

Since many participants of this conference are not experts in the domain of particle physics we will keep our explanations simple and we will start with a pedagogical introduction to the notion of a scattering cross section in CM and QM. We will also recall the proofs of OT in QM and in the relativistic *S* matrix theory. The plan of this paper is the following:

1. Scattering cross sections in CM and in QM.
2. Optical theorem in QM

3. Optical theorem in relativistic S-matrix approach.
4. Unitary *S* matrix without the optical theorem.
5. Data violating the optical theorem.
6. Conclusions

**2. Total and differential cross-sections**

The total cross section σ represents the effective interaction area of one beam and one target particle perpendicular to the beam of incoming particles. It depends in general on the particles involved, the energy of the beam etc. The immobile target can be replaced by another beam like in ISR or in LHC experiments For elastic scattering of two hard spheres with radiuses r and R: $\sigma = \pi (R+r)^2$.

If the interaction area σ is very small and if 1 cm$^2$ of a <u>thin</u> target contains *N* target particles then the probability *p* that one beam particle will be scattered by the target can be estimated as $p = N\sigma/1cm^2$. Since $p = (I_0 - I_{ns})/I_0$ where $I_0$ is a flux of incoming particles and $I_{ns}$ is a flux of non- scattered particles therefore

$$\sigma = \frac{N_s}{NI_0} \qquad (1)$$

where $N_s = (I_0 - I_{ns}) \times 1cm^2$ is a number of beam particles scattered by 1 cm$^2$ of the target per unit of time. Total cross-sections for hadron-hadron scattering are very small therefore they are measured in barns (b), mb, μb etc. where 1 b = 10$^{-28}$ m$^2$.

For long range interactions such as those described by a Coulomb potential total cross sections are infinite and a physical meaning have only so called differential cross section : $d\sigma(\Theta,\phi)$ geing the area such that a beam particle hitting it is scattered into the solid angle $d\Omega = sin\theta d\theta d\phi$.

$$d\sigma(\theta,\phi) = \frac{N_s(d\Omega)}{I_0 N} \qquad (2)$$

where $N_s(d\Omega)$ is a number of scattered beam particles in $d\Omega$ per unit time. Integrating $d\sigma(\Theta,\phi)$ over all angles one obtains the total cross section σ if it is finite.

In CM a scattering of one beam and one particle is described in CMS as a scattering of some fictitious point-like particle with reduced mass *m* on some immobile scattering center described by a potential *U(r)*. If the interaction has a cylindrical symmetry then $d\Omega = 2\pi sin\theta d\theta$ and one can find that $\sigma(\theta) = d\sigma(\theta) = 2\pi b db$ where $b = b(\theta)$ is the impact parameter of a beam particle . Using this functional relation for $U(r) = \alpha/r$ Rutherford obtained his famous formula:

$$d\sigma = \left(\frac{\alpha}{mv_\infty^2}\right)^2 \frac{d\Omega}{\sin^4\left(\frac{\theta}{2}\right)} \qquad (3)$$

and explained with success the scattering of a beam of alpha particles on a foil of gold.

**2. Scattering in QM and the optical theorem**

To describe the elastic scattering in QM one is solving a reduced relative motion wave equation in CMS [1-3]:

$$-\frac{\hbar^2}{2\mu}\Delta u - Vu = Eu \qquad (4)$$

with the asymptotic boundary condition for $r \to \infty$:

$$u \approx A\left(e^{ikz} + f(\theta)\frac{e^{ikr}}{r}\right) \quad (5)$$

Differential cross-section is calculated using the ratio of the outgoing and incoming probability density currents according to the formula:

$$\sigma(\theta) = \frac{J_r r^2}{J_i} = |f(\theta)|^2 \quad (6)$$

where $J = \text{Re}(\bar{\Psi}\frac{\hbar}{mi}\nabla\Psi)$.

Using the partial wave expansion of $u$ one obtains:

$$f(\theta) = (k)^{-1}\sum_{l=0}^{\infty}(2l+1)\sin\delta_l e^{i\delta_l} P_l(\cos\theta) \quad (7)$$

and the total cross-section:

$$\sigma = \frac{4\pi}{k^2}\sum_{l=0}^{\infty}(2l+1)\sin^2\delta_l \quad (8)$$

By comparing (7) and (8) one obtains immediately OT for the scattering of probability waves by short range potentials:

$$\sigma = \frac{4\pi}{k}\text{Im}(f0) \quad (9)$$

As we mentioned in the introduction OT is also proven for the scattering with various non-elastic channels. In high energy relativistic domain to describe the scattering of elementary particles one must abandon methods of nonrelativistic QM and use a relativistic $S$ matrix [4-6].

**3. Optical theorem in relativistic S matrix theory.**

When two hadrons collide various outcomes are possible and are called *channels of reaction:* 1+2→ 1+2, 1+2 →1+3+…+N or 1+2 →3+4+…+N. It is clearly more complicated than phenomena described by the scattering of probability waves on some potential.

In QFT and in relativistic $S$ matrix approach initial two particle state are represented by vectors $|i>$ in a Fock space and final states are represented by vectors $|f>$. The probability $P_{if}$ for obtaining a particular final state $|f>$ from the initial state $|i>$ is:

$$P_{if} = |\langle f|S|i\rangle|^2 \quad (10)$$

where $S$, called $S$ matrix, is a unitary operator.

The optical theorem is obtained by using a unitarity condition $SS^\dagger = S^\dagger S = I$ and decomposition:

$$S = I + iT \quad (11)$$

where $T$ is called a scattering operator.

From the unitarity condition $S^\dagger S = I$ using (11) one obtains $(I+iT)^\dagger (I+iT) = I$ and subsequently

$$-i<a|(T-T^{\dagger})|a> = <a|T^{\dagger}T|a> \qquad (12)$$

where $|a>$ is some initial two particle state vector.

Finally by placing a spectral decomposition of the identity operator between $T^{\dagger}$ and $T$ we obtain:

$$2\mathrm{Im}\langle a|T|a\rangle = \sum_b \langle a|T^{\dagger}|b\rangle\langle b|T|a\rangle = \sum_b |\langle b|T|a\rangle|^2 \qquad (13)$$

By interpreting $f_{ab} = <b|T|a>$ as a probability amplitude for obtaining a final state $|b>$ after the interaction of the particles in the state $|a>$ one recovers OT:

$$\mathrm{Im}\, f_{aa} = C(s)\sigma \qquad (14)$$

where $C(s)$ is some function of CMS energy of initial particles and $f_{aa}$ is a forward elastic scattering amplitude. For a binary reaction 1+2=3+4 the amplitude $f_{ab}$ depends on two kinematical relativistic variables $s=(p_1+p_2)^2=(p_3+p_4)^2$ and $t=(p_3-p_1)^2$ where $p_i$ are four-momenta of the particles. For the elastic scattering in CMS one finds $t=-p^2(1-cos\Theta)$ where $p$ is the modulus of the linear momentum of the first particle and $\Theta$ is its scattering angle. For the forward elastic scattering $t$ and $\Theta$ are equal to 0.

## 4. Unitary S matrix without the optical theorem

Various particle beams prepared in accelerators are manipulated using the classical relativistic mechanics and electrodynamics and projected on some targets. To describe the motion of free hadrons their internal degrees of freedom such as a quark structure are not important and a point-like approximation is completely justified when they are far apart.

We may therefore describe hadron-hadron scattering as a two-step process.

1. Using the information about prepared beam profiles and properties of the target we find the probabilities that two hadrons collide with a particular impact parameter $b$.

2. For $b$ smaller than the effective range of strong interactions we find probabilities for observing any particular final state $f$. These probabilities depend on $f$, on quantum numbers $\mu$ describing <u>all the relevant external and internal degrees of freedom</u> of the initial colliding pairs and of course on the detailed model for strong interactions.

Let us construct a mathematical model based on a unitary $S$ matrix having these properties [7-10]. First of all we split the quantum numbers $\mu$ into two sets such that there is a strong interaction only if $\mu \epsilon A$ and $|\mu> \epsilon H_2$. This splitting implies the splitting of the Hilbert space of all states into a direct sum of three Hilbert spaces:

$$H = H_1 \oplus H_2 \oplus H_3 \qquad (15)$$

where $H_1 = \{|\mu\rangle, \mu \notin A\}$, $H_2 = \{|\mu\rangle, \mu \in A\}$ and $H_3$ contains all other possible final states. Since pairs of particles are prepared with different impact parameters therefore the initial state for the scattering is described by a density operator:

$$\hat{D}_i = \sum_{\mu} \rho(\mu)|\mu\rangle\langle\mu| \qquad (16)$$

The sum in (15) and in equations which follow can be replaced by an integral over values of the impact parameter. According to our model a unitary S matrix can be written now in a form:

$$S = I \oplus \tilde{S} \qquad (17)$$

where $\tilde{S}$ is a unitary scattering operator $\tilde{S}\tilde{S}^\dagger = I$ acting from a subspace $H_2$ into a subspace of all possible final states. In our model (17) replaces (11) and one cannot prove OT.

Using (17) and the final density matrix $\hat{D}_f = S\hat{D}_i S^\dagger$ one finds the probability $P_{if} = Tr(|f\rangle\langle f| \hat{D}_f)$ for finding a final state $f$:

$$P_{if} = \sum_{\mu \notin A} \rho(\mu) |\langle f | \mu \rangle|^2 + \sum_{\mu \in A} \rho(\mu) |\langle f | \tilde{S} \mu \rangle|^2 \qquad (18)$$

For final states produced after strong interactions took place $\langle f | \mu \rangle$ vanish and only the second term in (18) corresponds to the scattering due to strong interactions:

$$P_{if} = \sum_{\mu \in A} \rho(\mu) |\langle f | \tilde{S} \mu \rangle|^2 \qquad (19)$$

## 5. Data violating the optical theorem

It is difficult to test OT in a reliable way since it requires the knowledge of non-measurable imaginary part of forward elastic scattering probability amplitude.

However using OT one may prove that a forward differential elastic cross section due to strong interactions for any spin state described by a density matrix $\rho$ has to satisfy the following inequality:

$$\left(\frac{d\sigma}{dt}\right)^\rho_{t=0} \geq \frac{(\sigma^\rho_{tot})^2}{16\pi\hbar^2} \qquad (20)$$

The direct test of (20) consists on estimating the value of elastic differential cross section in the forward direction (for $t=0$) and on comparing it with the data for total cross sections.

The main difficulty is the extrapolation to the region in which we do not have any data points. The fits to the data are done using some *reasonable* formulas containing free parameters on which the experimentalists reached a consensus. However using these *reasonable* formulas and the same set of data one can show [9, 27, 33-35] that OT is consistent with the data or that it is dramatically violated.

Our model has also other serious implications for any theoretical description of hadron-hadron scattering [9, 10]. We are assuming that initial two hadron states are mixed quantum states with respect to the impact parameter. The impact parameters cannot be controlled during the preparation of the beams and the targets therefore their statistical distribution may depend on the geometry of the beams and other factors.

In order to find such effects one may use the purity tests discussed by us in several papers [29-32]. One may find more details in our recent preprint [35]. These details together with other topics will be published in more technical paper addressed to elementary particle physicists..

## 6. Conclusions

We proved that one can describe short range hadron-hadron interactions by a unitary *S* matrix without being able to prove OT. Therefore OT is not a fundamental law of Nature and it could be violated in high energy hadron-hadron scattering.

The violation of OT would be confirmed if elastic scattering cross-sections at LHC and beyond were more suppressed than it was allowed by OT constraint. This constraint has been called *unitarity constraint* what was misleading since one can have a unitarity without OT [7, 8, 27, 36].

In LHC we have huge amount of inelastic scattering data with an average number of tracks observed bigger than 25. The main objective of LHC experiment is a search for Higgs particle, study of the production and the properties of charmed and other heavy flavor hadrons. In some sense the main objective is a search for a New Physics requiring the modification of SM [19].

Pursuing this goal one should not overlook another possibility for a New Physics namely: *Strong Interactions without the Optical Theorem* what if confirmed would be a major discovery.

To be able to make this discovery one has to reanalyze Tevatron, LHC and UHECR data without using OT constraint.

## Acknowledgments

I would like to thank Andrei Khrennikov for the invitation and a warm hospitality extended to me during QTAP conference. I would like also to thank UQO for a travel grant.